\documentclass[twocolumn,aps,prb]{revtex4-1}
\usepackage{graphicx}
\usepackage{epstopdf}
\usepackage{amssymb}
\usepackage{amsmath}

\usepackage{bm}
\usepackage{subfigure}
\usepackage{sidecap}

\newcommand{\be}{\begin{equation}}
\newcommand{\ee}{\end{equation}}

\newcommand{\bea}{\begin{eqnarray}}
\newcommand{\eea}{\end{eqnarray}}
\newcommand{\bd}{\begin{displaymath}}
\newcommand{\ed}{\end{displaymath}}
\newcommand{\ba}{\begin{array}}
\newcommand{\ea}{\end{array}}
\newcommand{\bi}{\begin{itemize}}
\newcommand{\ei}{\end{itemize}}
\newcommand{\bc}{\begin{center}}
\newcommand{\ec}{\end{center}}
\newcommand{\bfl}{\begin{flushleft}}
\newcommand{\efl}{\end{flushleft}}
\newcommand{\bfr}{\begin{flushright}}
\newcommand{\efr}{\end{flushright}}

\newcommand{\bl}{\begin{aligned}}
\newcommand{\el}{\end{aligned}}

  \def\bd{{\bf d}}

\def\6{\partial}

\def\={\!\!\!&=&\!\!\!}
\def\+{\!\!\!&&\!\!\!+~}
\def\-{\!\!\!&&\!\!\!-~}

\def\st{\sin\theta}
\def\ct{\cos\theta}
\newcommand{\vc}[1]{\boldsymbol{#1}}
\begin{document}
\date{\today}
\title{Magnetic Excitations in Spin-Orbit Coupled $d^4$ Mott Insulator on 
Square Lattice }

\author{Alireza Akbari and Giniyat Khaliullin}
\affiliation{Max-Planck-Institut f\"ur Festk\"orperforschung, 
Heisenbergstrasse 1, D-70569 Stuttgart, Germany} 

\begin{abstract}
We study the magnetic order and excitations in strong spin-orbit coupled, 
Van Vleck-type, $d^4$ Mott insulators on a square lattice. Extending the 
previous work, we include the tetragonal crystal field splitting and explore 
its effects on magnetic phase diagram and magnon spectra. Two different ordered 
phases, with in-plane and out-of-plane orientation of the staggered moments, 
are found for the higher and lower values of the crystal field splitting,
respectively. The magnetic excitation spectra for paramagnetic and 
magnetically ordered phases are calculated and discussed in the context 
of a candidate spin-orbit $d^4$ Mott insulator Ca$_2$RuO$_4$.
\end{abstract}

\maketitle

\section{Introduction}

In a solid, five-fold orbital degeneracy of a $d$-electron level is lifted by 
crystal field potential as well as by covalency effects. In case of local cubic 
symmetry, two subsets of $d$-orbitals with two-fold $e_g$ and three-fold 
$t_{2g}$ symmetry, separated by a large energy of the order of 
$10Dq\sim 2-3$~eV, are formed. The remaining degeneracy of orbitals -- which 
adds up to that of spin -- has to be lifted one way or another, involving 
dynamical Jahn-Teller effect and interionic exchange interactions. If the 
latter mechanism dominates, the spin and orbital degrees of freedom strongly 
couple to each other and are described by a family of so-called 
Kugel-Khomskii-type models~\cite{Kug82}. 

General behavior of Kugel-Khomskii Hamiltionans is very complex because of the 
frustrated nature of orbital interactions, in particular in the case of 
$t_{2g}$ orbitals where the higher degeneracy enhances quantum 
effects~\cite{Kha00,Kha05}. In addition to that, $t_{2g}$ triplet has an 
unquenched orbital angular momentum $L$, and relativistic spin-orbit 
coupling (SOC) $\lambda(\vc{S}\cdot\vc{L})$ is active. When $\lambda$ is 
comparable to the strength of exchange interactions, spin-orbit coupling 
effects become of a nonperturbative nature. In that case, it is more 
convenient to represent the spin-orbital exchange Hamiltionans in terms of 
ionic multiplets in which SOC is already included~\cite{Ell68}. Often, it is 
sufficient to keep the lowest-lying ionic multiplet with $2\tilde{S}+1$ 
degeneracy; this results in effective, ''pseudospin $\tilde{S}$'' 
Hamiltonians describing low-energy magnetic properties of a material. 

By construction, pseudospins $\tilde{S}$ inherit the spatial shape and 
bond-directional nature of orbitals and their interactions~\cite{Kha05}. 
Thus, the pseudospin Hamiltonians may strongly deviate from a conventional, 
spin-isotropic Heisenberg models, even in a simplest case of just twofold 
Kramers degeneracy with $\tilde{S}=1/2$. As an example, exchange 
interactions between $t_{2g}^5$ ions with pseudospins $\tilde{S}=1/2$ obtain 
large Ising term with an unusual, bond-dependent alternation of the 
''Ising-axes''~\cite{Kha05,Jac09,Cha10}, leading to unconventional 
magnetic states. The pseudospin Hamiltonians for $\tilde{S}>1/2$ receive 
in addition strong biquadratic and multipolar interactions~\cite{Per09,Che11}. 
Experimental studies of the iridium oxides hosting pseudospin 
physics~\cite{Kim08,Kim09,Kim12} have boosted general interest in strong 
spin-orbit coupled magnetism (see Ref.~\onlinecite{Wit13} for the 
recent review). 

It may happen that the lowest spin-orbit ionic state has no degeneracy, 
$\tilde{S}=0$, and hence it is nonmagnetic. Such is the case of transition 
metal ions with $t_{2g}^4$ configuration, where the spin $S=1$ and orbital 
$L=1$ moments form a singlet ground state~\cite{Abr70}. Compounds with such 
nominally nonmagnetic (''Van Vleck-type'') ions may still undergo magnetic 
transitions, due to mixing of the ground state $\tilde{S}=0$ level with 
higher-lying $\tilde{S}=1,2$ multiplets by virtue of intersite exchange 
interactions~\cite{Kha13}. Because of SOC, the transitions between multiplets 
with different $\tilde{S}$ are magnetically active. In a solid, they become 
dispersive bands and have been observed in cobalt~\cite{Hol71,Buy71} and 
iridium~\cite{Kim12} oxides. Magnetic order in systems with $\tilde{S}=0$ can 
be thus viewed as a Bose condensation of excitonic $\tilde{S}=1$ band. 
A hallmark of such magnetism is the presence of soft amplitude 
mode~\cite{Gia08}, corresponding here to the length fluctuations of the 
total angular momentum $\tilde{\vc S}=\vc{S}+\vc{L}$, in addition 
to conventional spin waves. 

Theory of the exchange interactions and excitonic magnetism in Van Vleck-type 
$t_{2g}^4$ systems has been developed recently in Ref.~\onlinecite{Kha13}, and 
ruthenium oxide Ca$_2$RuO$_4$ was suggested as a possible candidate material, 
based on the experimental observation~\cite{Miz01} of an unquenched SOC in 
this compound. In this paper, we consider magnetic order and excitations in 
more detail, with a particular focus on the effects of tetragonal distortion 
generally present in most perovskites.  

\section{Model Hamiltonian}

\begin{figure}
\centerline{
(a)
\includegraphics[width=0.23\linewidth]{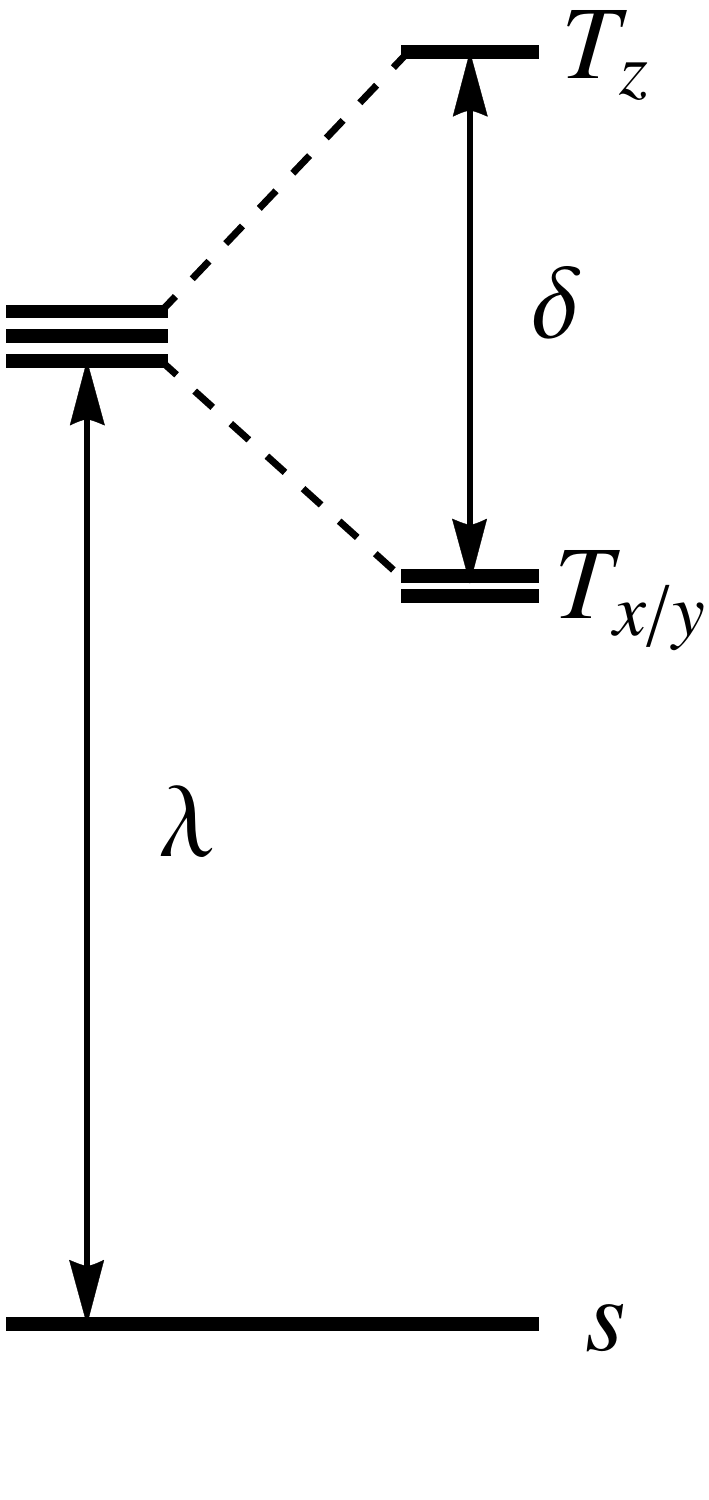}
\hspace{0.15cm} 
(b) \includegraphics[width=0.5\linewidth]{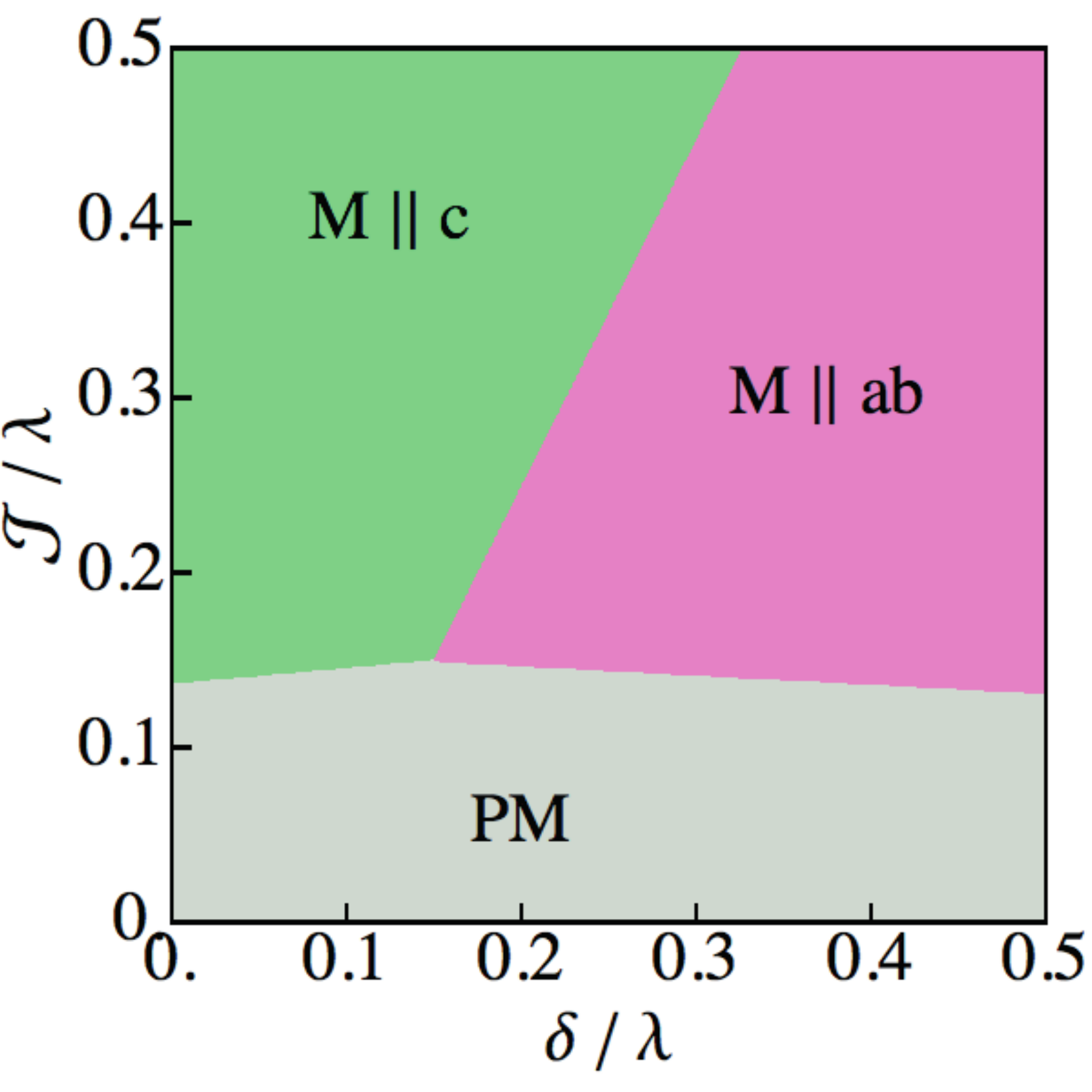}
}
\caption{(Color online) 
(a) Schematic of spin-orbit level structure including ground state singlet
and a higher-lying pseudospin $\tilde{S}=1$ triplet; the latter is split 
by tetragonal crystal field $\delta$ which favors in-plane orientation of
magnetic moments. 
(b) Magnetic phase diagram as a function of exchange coupling and crystal-field
splitting. It includes paramagnetic (PM) and two magnetic states with 
easy-axis ($M\parallel c$) and easy-plane ($M\parallel ab$) orderings.
} 
\label{PhaseFig}
\end{figure}

Having in mind a layered perovskite structure of Ca$_2$RuO$_4$, we consider
square lattice of $t_{2g}^4$ ions which are assumed to have a low-spin
configuration with spin $S=1$ and orbital $L=1$ moments. Intraionic SOC
generates three levels at energies $0, \lambda$, and $3\lambda$, corresponding 
to the spin-orbit multiplets with total angular momentum $\tilde{S}=0,1$, and 
$2$. We neglect the highest, $\tilde{S}=2$ multiplet at energy $3\lambda$; 
this is justified if the exchange interactions are not too strong as compared
to SOC parameter $\lambda$. The remaining ionic degrees of freedom include 
ground state singlet $|s\rangle$ and $\tilde{S}=1$ triplet 
$|T_{0,\pm1}\rangle$ states. In a $|M_S,M_L\rangle$ basis, the wave-functions 
read as $|s\rangle=\frac{1}{\sqrt 3}(|1,-1\rangle-|0,0\rangle+|-1,1\rangle)$, 
$|T_0\rangle=\frac{1}{\sqrt 2}(|1,-1\rangle-|-1,1\rangle)$,
$|T_{\pm 1}\rangle=\pm\frac{1}{\sqrt 2}(|\pm 1,0\rangle-|0,\pm 1\rangle)$. 
It calculations, the Cartesian components  
$T_x=\frac{1}{i\sqrt 2}(T_1-T_{-1})$, $T_y=\frac{1}{\sqrt 2}(T_1+T_{-1})$,
and $T_z=iT_0$ are often more convenient. Tetragonal crystal field splits 
the triplet level as shown Fig.~\ref{PhaseFig}(a). We note that a positive 
$\delta>0$ corresponds to compression of the octahedra along $c$ axis 
(in a view of point-charge model), and its value is equal to the half of 
the tetragonal splitting between $xy$ and $xz/yz$ orbital levels, 
$\delta=(E_{xz/yz}-E_{xy})/2$, which would be expected in the limit 
of $\lambda=0$.  

The effective singlet-triplet model ${\cal H}_{eff}$ that we consider below 
reads then as follows:
\be
{\cal H}_{eff}={\cal H}_{CF}
+\lambda\sum_in_i+{\cal J}\! \sum_{<ij>_{a}}h_{ij}^{(a)}+{\cal J}\!
\sum_{<ij>_{b}}h_{ij}^{(b)},  
\label{T-S_Hamiltonian}
\ee 
comprising a tetragonal crystal field contribution 
\be
{\cal H}_{CF}=\delta
\sum_i{\Big (}
n_{iz}-\frac{1}{3}n_i 
{\Big )},
\ee
a spin-orbit coupling energy $\lambda$ of triplet states [second term in 
Eq.~(\ref{T-S_Hamiltonian})], and, finally, the superexchange interactions 
[last two terms in Eq.~(\ref{T-S_Hamiltonian})]. For the $a$-type bonds of
square lattice, the interactions can be represented via the hard-core 
$T$-bosons as follows~\cite{note1}: 
\be
 h_{ij}^{(a)}=\! 
   \vc T_{i}^{\dagger} \cdot \vc T_{j}^{}
   -\frac{1}{3} T_{i,x}^{\dagger}T_{j,x}^{}
   -\frac{5}{6} \vc T_{i}^{}\cdot \vc T_{j}^{}  
   +\frac{1}{6} T_{i,x}^{}T_{j,x}^{}
  +h.c.
\label{exchange} 
\ee
Interactions $h_{ij}^{(b)}$ on $b$-bonds are obtained by a substitution  
$T_{i,x} \rightarrow T_{i,y}$. In the above equations, $n=n_x+n_y+n_z$ with 
$n_\alpha=T^\dagger_\alpha T_\alpha$, and ${\cal J} =\frac{t_0^2}{U}$
represents the exchange energy scale. Note that $h^{(a,b)}$ represent 
the quadratic terms in $T$-boson interactions; full exchange Hamiltonian 
contains also three- and four-boson terms~\cite{note2} which are neglected 
here. This approximation is valid near the critical points when density of
condensed bosons is small. We note finally that $T$ operators physically 
correspond to the singlet-triplet transitions between $\tilde{S}=0$ and 
$\tilde{S}=1$ levels. In other words, they are composite objects subject 
to the hard-core constraint $n_T\leq 1$, and can alternatively be 
represented as $\vc T\rightarrow s^{\dagger}\vc t$, via singlet $s$ and
triplet $\vc t$ particles that obey the constraint $n_s + n_t=1$. 

\section{Ground state properties}

Depending on the relative strength of the exchange ${\cal J}$ and SOC $\lambda$
parameters, the ground state of effective Hamiltonian 
Eq.~(\ref{T-S_Hamiltonian}) can be either paramagnetic or antiferromagnetic. 
There are two different magnetic phases, with out-of-plane $M\parallel c$ 
and in-plane $M\parallel ab$ orientations of the staggered moments. 
The $M$-orientation is decided by the competition between the exchange 
${\cal J}$ and the crystal-field $\delta$ couplings. We calculate below 
classical energies of magnetically ordered states, and obtain from them 
a phase diagram and ordered moment values.  

\begin{figure}
\centerline{
\includegraphics[width=0.7\linewidth]{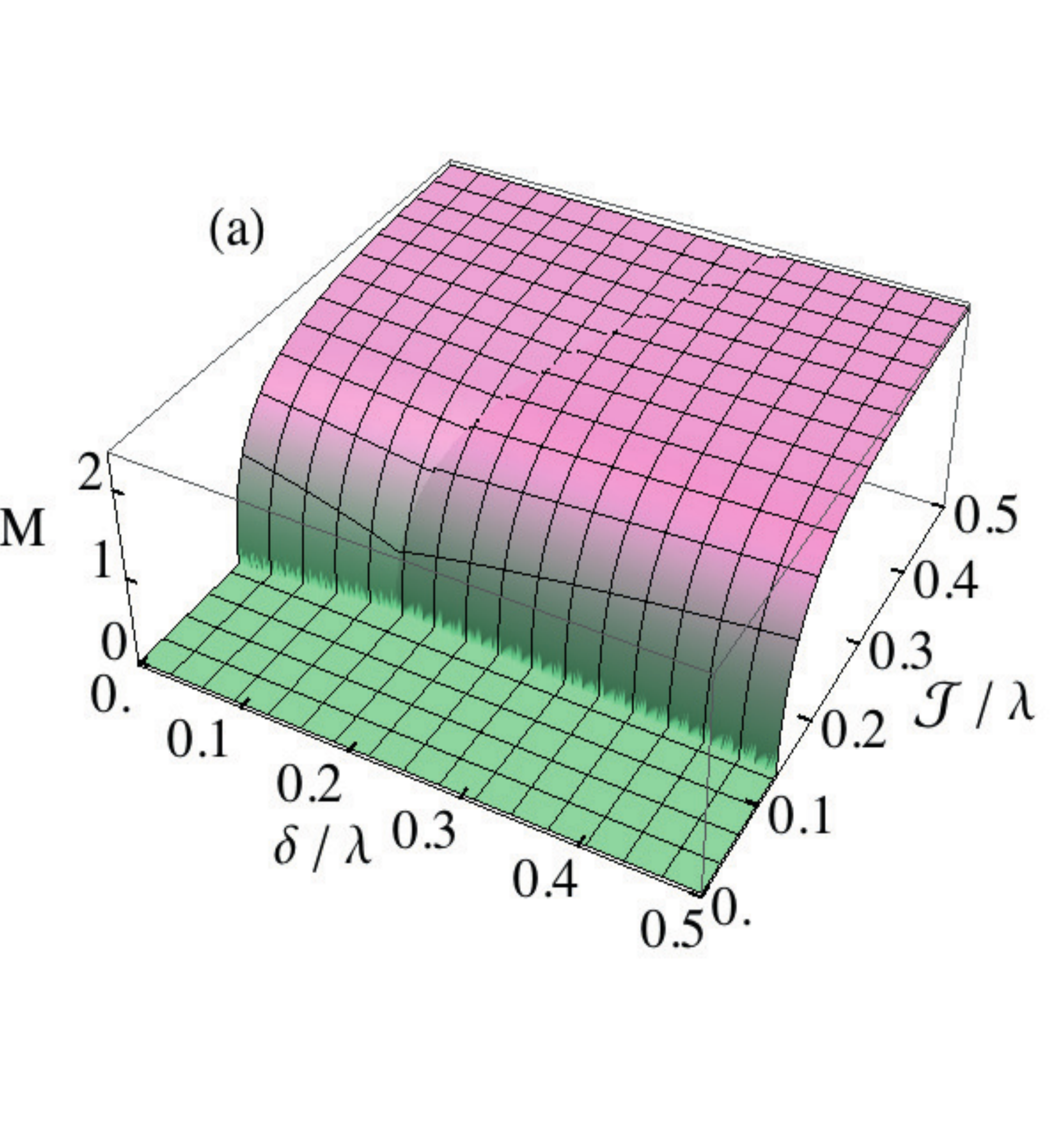}
}\vspace{-1.0cm}
\centerline{
\includegraphics[width=0.98\linewidth]{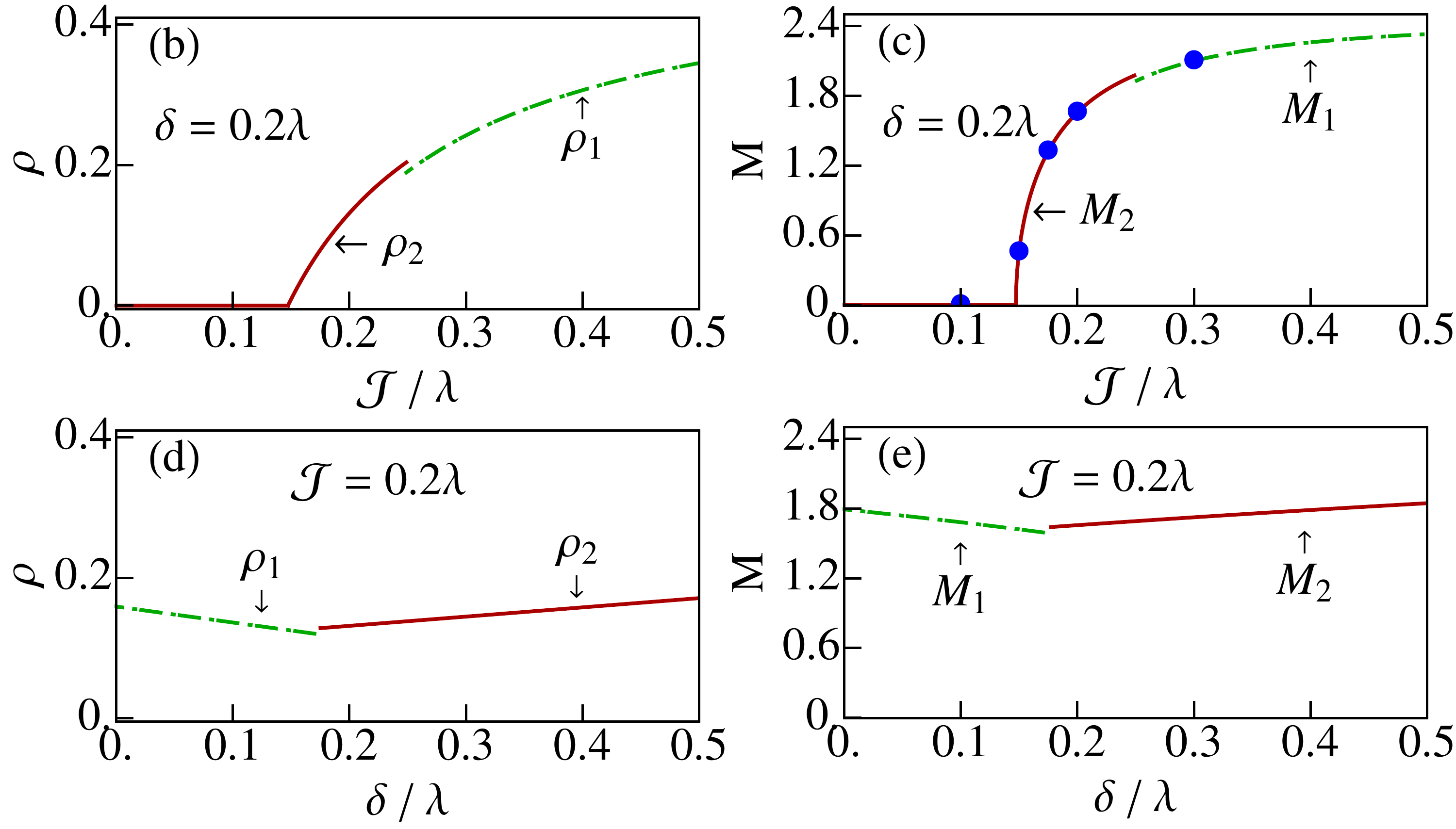}
}
\caption{(Color online) 
(a) The staggered magnetic moment $M$ as a function of $\delta$ and ${\cal J}$. 
(b) The condensate density $\rho$ and (c) the staggered moment $M$ 
values versus exchange coupling ${\cal J}$, for the fixed crystal-field
splitting $\delta=0.2\lambda$. [The blue dots in panel (c) will be referred 
to later in Fig.~\ref{Figsusceptibility}].
(d) The condensate density $\rho$ and e) the staggered moment $M$ 
values versus $\delta$, for the fixed ${\cal J}=0.2\lambda$. 
} 
\label{MagnetizationFig}
\end{figure}

\subsection{Phase diagram}
{\bf Magnetic phase I} (${M  \parallel c}$): This state is obtained by  
a condensation of the $T_z$-component of $\tilde{S}=1$ triplet: 
$T_z\rightarrow \sqrt\rho_{1}$. The corresponding classical energy gain is:  
\be
E_{g1}=-\rho_{1}\mu_{1}
=-\frac{1}{4\kappa_{1}}
(\kappa_{1}^{ }-\beta_{1}^{ })^{2 },
\ee
where $\rho_{1}=\frac{1}{2}(1-\frac{\beta_{1}}{\kappa_{1}})$ is the condensate
density, and $\mu_{1}= \frac{1}{2}(\kappa_{1} -\beta_{1})$. We note that the 
constant $-\mu$ has a physical meaning of chemical potential. The condensate 
density and hence the magnetic moments are determined by the interaction 
parameters $\kappa_{1}=\frac{22}{3}{\cal J}$ and 
$\beta_{1}=\lambda+\frac{2}{3}\delta$. 
The magnetic phase transition sets in at $\kappa_{1}=\beta_{1}$ (i.e., when
$\mu_{1}$ becomes zero); this gives the critical value of the exchange 
constant ${\cal J}$ as ${\cal J}_{1}^c=(3\lambda+2\delta)/22$.\\

{\bf Magnetic phase II  } (${M  \parallel ab}$):
The magnetic moment is in the $ab$-plane corresponding to the condensation of 
$T_x\rightarrow \sqrt\rho_{2}$. The ground state energy can be represented in 
the form as above,  
\be
E_{g2}=-\rho_{2}\mu_{2}
=-\frac{1}{4\kappa_{2}}
(\kappa_{2}^{ }-\beta_{2}^{ })^{ 2},
\ee
but with different parameters: 
$\rho_{2}=\frac{1}{2}(1-\frac{\beta_{2}}{\kappa_{2}})$ and 
$\mu_{2}= \frac{1}{2}(\kappa_{2} -\beta_{2})$, where
$\kappa_{2}=\frac{19}{3}{\cal J}$ and $\beta_{2}=\lambda - \frac{1}{3}\delta$. 
The magnetic phase transition line is given by $\kappa_{2}=\beta_{2}$,   
and the critical value ${\cal J}^{c}_{ 2}=(3\lambda-\delta)/19$.

Using the above results, we find a phase diagram as shown in
Fig.~\ref{PhaseFig}(b). At small crystal fields $\delta$, the exchange 
anisotropy terms in Eq.~(\ref{exchange}) select out-of-plane $M$-direction. 
However, already quite small tetragonal splitting $\delta$ stabilizes 
the in-plane magnetic order, which corresponds to the case of Ca$_2$RuO$_4$. 

\subsection{Staggered Magnetisation }

The magnetic moment of present singlet-triplet system is represented by 
the following operator~\cite{Kha13}:
\be
\vc M=-i \sqrt{6}({\vc T}-{\vc T}^\dagger)-
i g_J({\vc T}^\dagger \times {\vc T}), 
\ee
with $g_J=1/2$. In magnetic phases with condensed bosons, the first term of 
this operator obtains finite expectation value [at the ordering wave-vector 
$(\pi,\pi)$]. Using the above results for condensate amplitudes, we find that 
the staggered magnetic moment in phase I is:    
\be
M_{1}=\sqrt{6(1-\eta_{1})} \; ;                       {\text\rm \;\;  \;\; }
\;\;{\cal J}>{\cal J}_{1}^c \; ,
\label{SMagnetization}
\ee
where $\eta_{1}=\beta_{1}^2/\kappa_{1}^2$. The same equation holds for the 
magnetic moment in phase II, but with $\eta_{2}=\beta_{2}^2/\kappa_{2}^2$ 
and ${\cal J}>{\cal J}_{2}^c$. Parameters $\beta_{1,2}$ and $\kappa_{1,2}$
have been given above.  

The numerical results for the staggered moment as a function of parameters 
$\delta$ and ${\cal J}$ are shown in Fig.~\ref{MagnetizationFig}. 
A clear trace of the phase transition from PM to magnetic phases, and 
a discontinuous spin-reorientation transition between phases I and II are 
observed. 

Condensate densities [Fig.~\ref{MagnetizationFig}(b)] and staggered moments  
[Fig.~\ref{MagnetizationFig}(c)] critically depend on ${\cal J}/\lambda$
ratio. However, they are not sensitive to the value of anisotropy parameter
$\delta$ [Fig.~\ref{MagnetizationFig}(d,e)] whose major effect is the
stabilization of phase II with in-plane magnetic moments. 

\section{Excitation Spectra}
\subsection{Magnon dispersions}
We consider now spin excitations above the ground state. Technically, we 
follow early works~\cite{Som01,Mat04} which extended a linear spin-wave
theory to singlet-triplet models. We handle the particle-number 
constraint on average only, neglecting magnon interaction effects. 

Within this approximation, spin excitations in the paramagnetic phase 
follow directly from Eqs.~(\ref{T-S_Hamiltonian}-\ref{exchange}),
after the Bogoliubov transformation of the $T$ operators (in momentum space). 
For $T_z$ component, this gives 
\be
\omega_z(\vc k)=(\lambda+\frac{2}{3}\delta) 
\sqrt{1+a_z\phi_{\vc k}}\;, \;a_z=\frac{22{\cal J}}{3\lambda+2\delta}\;, 
\ee
where $\phi_{\vc k}=\frac{1}{2} (\cos{k_{x}}+\cos{k_{y}})$ is a square lattice 
form-factor. Because of tetragonal symmetry, the $T_x$ and $T_y$ modes 
are degenerate:  
\be
\omega_{x/y}(\vc k)=(\lambda-\frac{1}{3}\delta) 
\sqrt{1+a_{x/y}\phi_{\vc k}}\;, \;a_{x/y}=\frac{19{\cal J}}{3\lambda-\delta}\;. 
\ee

For the antiferromagnetically ordered phases, we introduce two sublattices 
labeled by A and B. It is convenient also to introduce the sublattice 
dependent phase shifts $T_{\rm  A}\rightarrow iT$, $T_{\rm  B}\rightarrow
-iT$, and work within the extended Brillouin zone (BZ). Then, after the  
Fourier transformation ${\vc T}_k=\sum_i  e^{-i{\vc k}\cdot{{\vc r}_i}} {\vc T}_i$ in 
Eqs.~(\ref{T-S_Hamiltonian}-\ref{exchange}), we arrive at the following
momentum-space Hamiltonian:
\bea
&&
\nonumber
{\cal H}= \sum_{k}
( {\cal H}_{k}^z+{\cal H}_{k}^x+{\cal H}_{k}^y)
\\
&&
\hspace{0.2cm}
=
\nonumber
\!\sum_{k}
(\lambda+\!\frac{2}{3}\delta-\!4{\cal J}\phi_{\vc k})  
   T_{\vc k, z}^{\dagger} T_{\vc k, z}^{}
-\frac{5}{3} {\cal J}\phi_{\vc k}
  (T_{\vc k, z}^{} T_{\vc k, z}^{}+\!h.c.) 
  \\\nonumber
&&\hspace{0.2cm}  
   + {\Big[}  
(\lambda-\frac{1}{3}\delta +\!\frac{10}{3}  {\cal J}\!\phi_{\vc k})
T_{\vc k, x}^{\dagger} T_{\vc k, x}^{}
-\frac{3}{2}  {\cal J}\phi_{\vc k}
(T_{\vc k, x}^{} T_{\vc k, x}^{}+\!h.c.)  
  {\Big]}
   \\
&&\hspace{0.2cm}  
   +
     {\Big [}
x\rightarrow y
  {\Big]},
\label{Hk}
\eea

Magnetic order in singlet-triplet models implies condensation of a particular 
component of the triplet state, i.e., it mixes-up coherently with the ground 
state singlet. In order to describe this process, we introduce 
${\vc T}\rightarrow s^{\dagger}{\vc t}$ with $n_s+n_t=1$, and transform the
basis as follows: 
\be
\bl
t_{\alpha}=\tilde{s}\st + \tilde{t}_{\alpha}\ct, \\
         s=\tilde{s}\ct - \tilde{t}_{\alpha}\st,
\el
\ee
where $\alpha=z(x)$ for phase I (II). A new $\tilde{s}$ boson is then condensed.
Fluctuations of $\tilde{t}_{\alpha}$ represent amplitude fluctuations, while
remaining two (uncondensed) components of the triplet become transverse 
magnons. The basis-rotation angle $\theta$ is determined by minimization of the
classical energy $E_g$ of Hamiltonian Eq.~(\ref{Hk}), which results in 
$\st=\sqrt{\rho}$ and $E_g = -\rho\mu$, with condensate densities
$\rho_{1,2}$ and potentials $-\mu_{1,2}$ for phases I and II, correspondingly,
as given in a previous section. 

We consider first the magnetic phase I. After the above transformations, 
quadratic part of the Hamiltonian Eq.~(\ref{Hk}) takes the following form:
\be
{\cal H}_{k}^{z}={\cal A}_{k}^{z} \;  
\tilde{t}_{\vc k,z}^{\dagger} \tilde{t}_{\vc k,z}^{}+
\frac{1}{2}{\cal B}_{k}^{z} 
{\Big(} \tilde{t}_{\vc k,z}^{} \tilde{t}_{-\vc k,z}^{}+ h.c.  
{\Big)},
\label{z}
\ee
where
\be
\bl
&
{\cal A}_{  k}^{z}=\kappa_{1} {\Big[} 
1+\frac{\phi_{\vc k}}{22}(1+11\eta_{1}) {\Big]}, \\
&
{\cal B}_{  k}^{z}=\kappa_{1} 
{\Big[} 
\frac{\phi_{\vc k}}{22}(1-11\eta_{1}){\Big]}.
\el
\ee
Diagonalization of Eq.~(\ref{z}) gives the amplitude mode dispersion:  
\be
\omega_z(\vc k)=\sqrt{({\cal A}_k^z-{\cal B}_k^z)
({\cal A}_k^z+{\cal B}_k^z)}\; 
= \sqrt{\kappa_{1}^2+ \beta_{1}^2 \phi_{\vc k}}\;.
\ee
The transverse components $t_{x/y}$ are degenerate in phase I. Accounting for
the chemical energy shift $-\mu(n_x+n_y)$, we find the corresponding quadratic
Hamiltonian for $x/y$ modes in a form of Eq.~(\ref{z}) again, with the 
following constants 
\be
\bl
&
{\cal A}_k^x={\cal A}_k^y=
{\bar \kappa}_{1} -\delta+
\frac{5}{11}{\bar \kappa}_{1}\phi_{\vc k},
  \\
  &
  {\cal B}_{  k}^{x}={\cal B}_{  k}^{y}=
\frac{9}{22}{\bar \kappa}_{1}\phi_{\vc k}, 
\el
\ee
where ${\bar \kappa_{1}}=(\kappa_{1}+\beta_{1})/2$. This gives spin-wave
dispersions 
\be
\bl
&
\omega_{x/y}(\vc k)=({\bar \kappa_{1}}-\delta)
\sqrt{1+\frac{19}{22}
\frac{{\bar \kappa_{1}} \phi_{\vc k}} {({\bar \kappa_{1}}-\delta)}}\;
\el
\ee
for the magnetic phase I with $ M\parallel c$. 

\begin{figure}[t]
\centerline{
\includegraphics[width=0.95\linewidth]{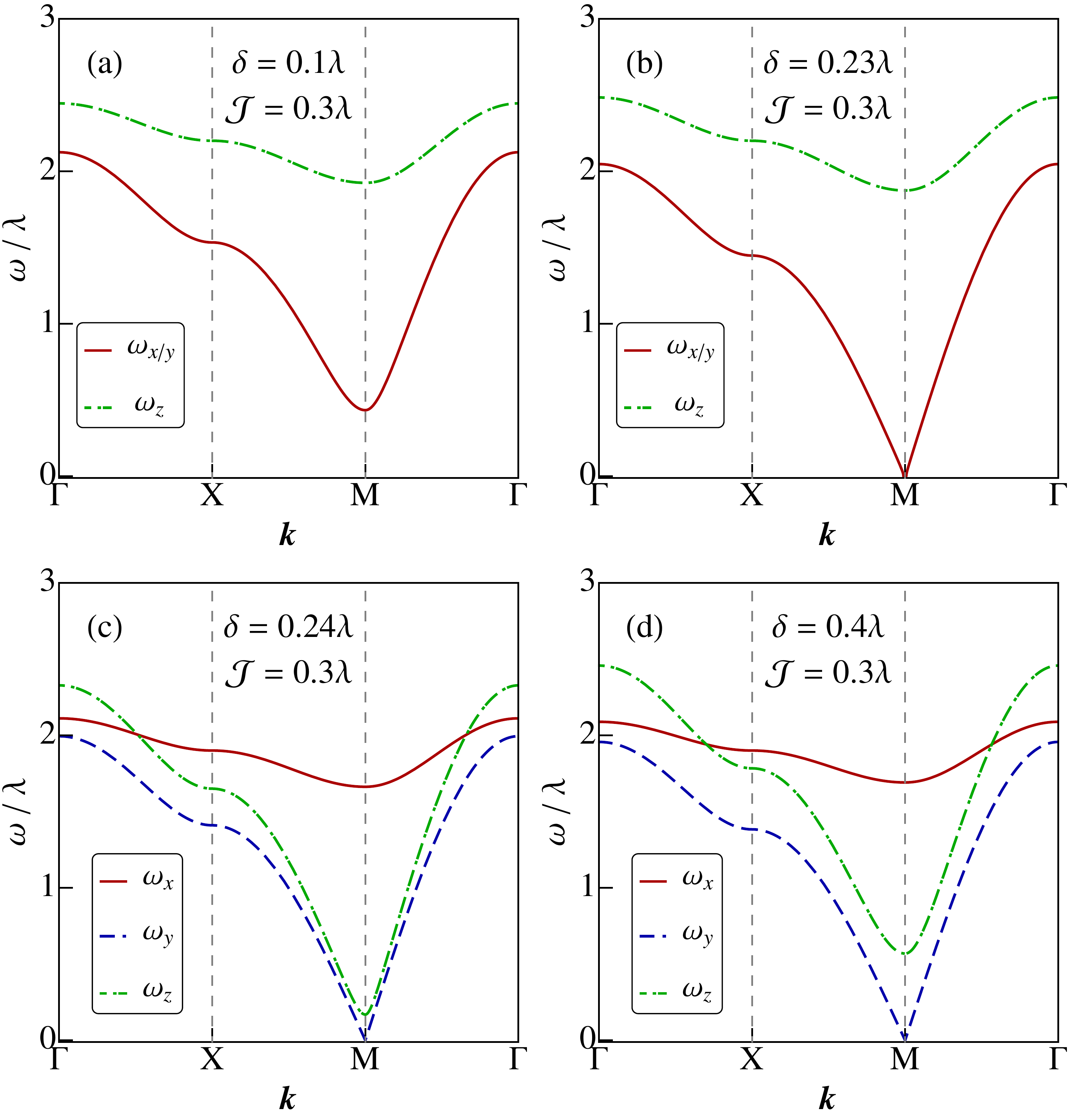}
}
\caption{(Color online) 
The excitation energies $\omega_\alpha$ (in units of SOC constant $\lambda$) 
versus momentum $\vc k$, for fixed ${\cal J}=0.3\lambda$ and
different $\delta$ values:  
(a) $\delta=0.1\lambda$, (b) $\delta=0.23\lambda$, (c) $\delta=0.24\lambda$,
and (d) $\delta=0.3\lambda$. 
Panels (a-b) represent the phase I, and (c-d) represent the phase II. 
Here, $\Gamma=(0,0)$, $X=(0,\pi)$ and $M=(\pi,\pi)$. 
} 
\label{OmegaVsK_delta}
\end{figure}

\begin{figure}[t]
\centerline{
\includegraphics[width=0.95\linewidth]{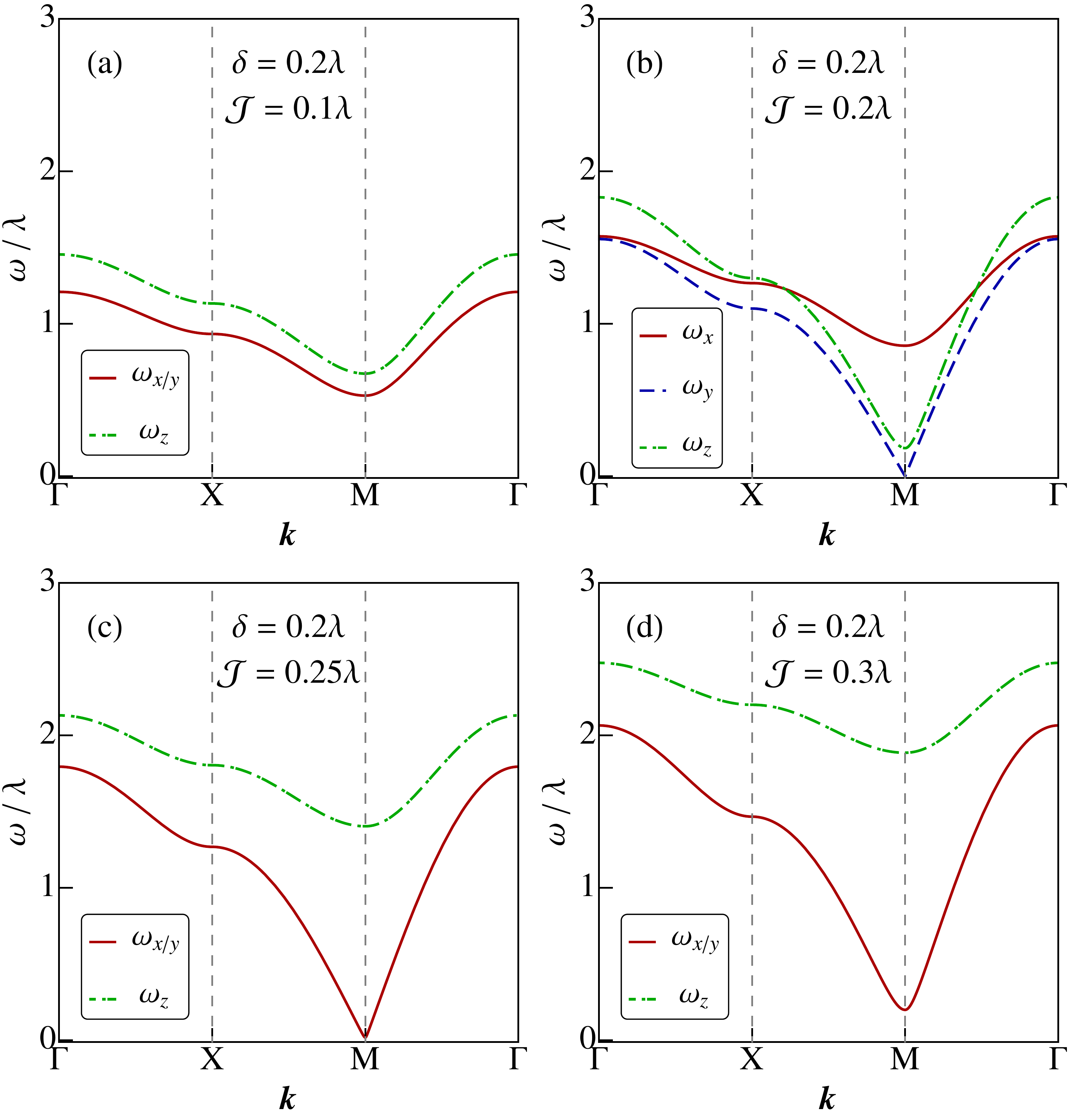}
}
\caption{(Color online) 
The same as the Fig.~\ref{OmegaVsK_delta} but for fixed $\delta=0.2\lambda$
and different ${\cal J}$ values: (a) ${\cal J}=0.1\lambda$, 
(b) ${\cal J}=0.2\lambda$, (c) ${\cal J}=0.25\lambda$, 
and (d) ${\cal J}=0.3\lambda$. 
Panel (a) corresponds to the paramagnetic phase, (b) represents the phase II,
and (c-d) represent the phase I.
} 
\label{OmegaVsK_J}
\end{figure}

For the magnetic phase II, similar calculations give the following results for 
the energy-momentum dispersions of the amplitude ($x$) and transverse ($y,z$) 
modes: 
\bea
\bl
&
\omega_{x}(\vc k)=
\sqrt{ \kappa_{2}^2 +
 \beta_{2}^2 \phi_{\vc k} 
  }
  \;,
 \\
 &
\omega_{y}(\vc k)=
 {\bar \kappa_{2}} 
 \sqrt{1+\phi_{\vc k}}
\;,
\\
&
\omega_{z}(\vc k)=
({\bar \kappa_{2}}+\delta)
\sqrt{1+
\frac{22}{19}
\frac{
{\bar \kappa_{2}} \phi_{\vc k}} {({\bar \kappa_{2}}+\delta)}
}
\;.
\el
\eea
It is noticed that in phase II with $ M\parallel ab$, there is no degeneracy 
of magnon branches, i.e., in-plane ($y$) and out-of-plane ($z$) magnons 
are split. 

Some examples of magnon dispersion curves, representing different magnetic 
phases, are plotted in Figs.~\ref{OmegaVsK_delta} and~\ref{OmegaVsK_J}. 
Fig.~\ref{OmegaVsK_delta} shows the evolution of excitation spectra as 
a function of the crystal-field parameter $\delta$, and their dependence 
on the exchange parameter ${\cal J}$ is illustrated in Fig.~\ref{OmegaVsK_J}. 
The features mentioned above such as a separation of the amplitude mode from 
the low-energy magnon modes, and splitting of the latter into two distinct 
branches in phase II can be noticed. 

\begin{figure}
\centerline{
\includegraphics[width=0.98\linewidth]{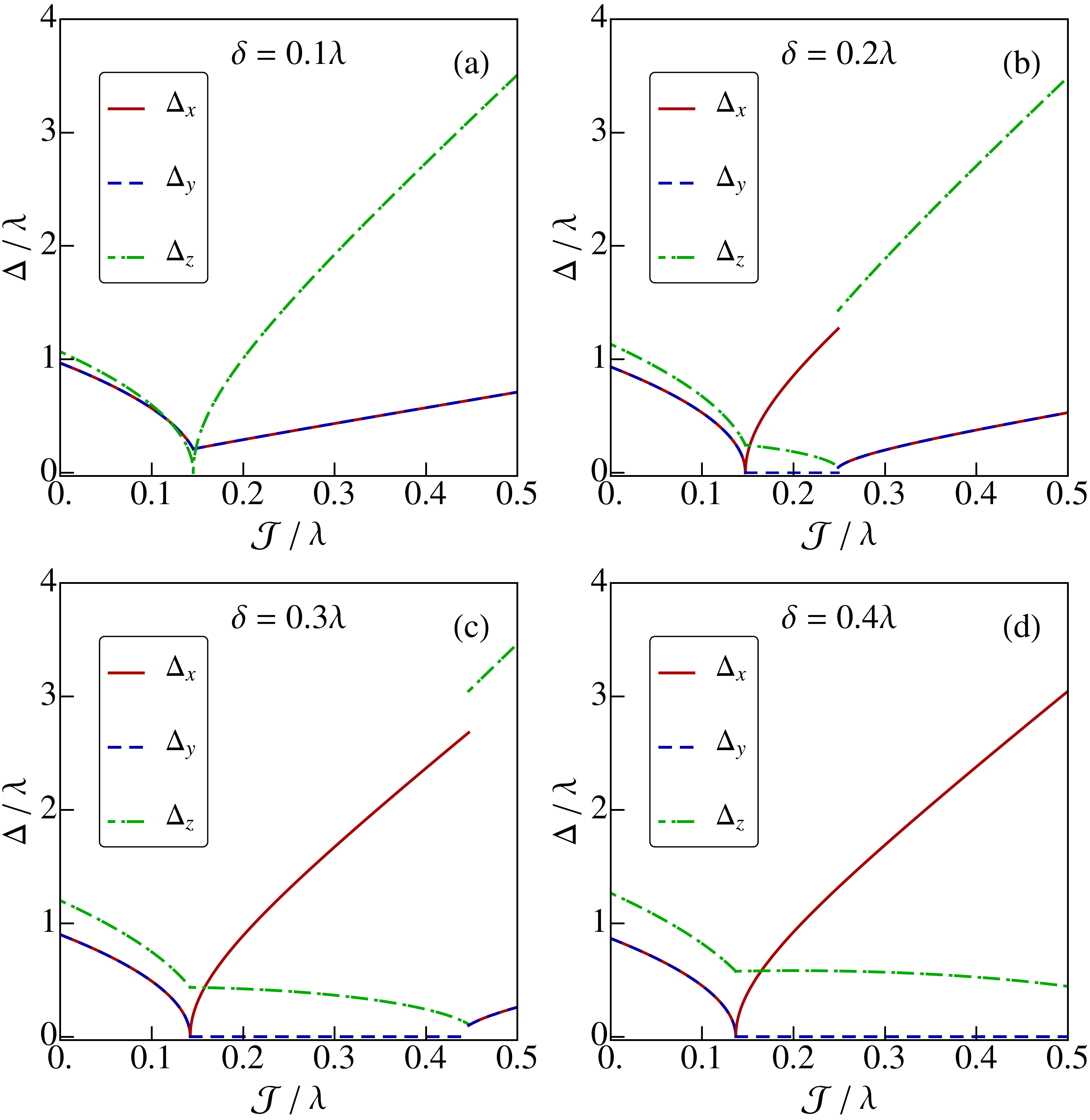}
}
\caption{(Color online) 
The excitation energy gaps at the Bragg ($M$) point, 
$\Delta=\omega(\vc Q)$, as a function of exchange interaction ${\cal J}$, 
for different values of crystal-field parameter: (a) $\delta=0.1\lambda$, 
(b) $\delta=0.2\lambda$, (c) $\delta=0.3\lambda$, (d) $\delta=0.4\lambda$. 
} 
\label{OmegaVsJ}
\end{figure}

In order to see the evolution of the magnon gaps in more detail, we plot in 
Fig.~\ref{OmegaVsJ} the magnetic excitation energies at the Bragg point 
$\vc Q= (\pi,\pi)$, as a function of the exchange constant ${\cal J}$ at 
different $\delta$ values. In the PM phase (small ${\cal J}$), all the 
branches have a finite gap. At the critical value of ${\cal J}$, gap for the
amplitude mode closes. Further increase of ${\cal J}$ enhances the excitation 
gaps for all the branches in phase I [see panel (a)]. In phase II, there 
remains gapless Goldstone mode [see panels (b-d)], corresponding to a free
rotation of the staggered moment within $ab$ plane. Figs.~\ref{OmegaVsJ}(b)
and (c) illustrate a transformation of spin-wave dispersions at the 
first order phase transitions between phases I and II. 

\subsection{Magnon Intensities}

The intensity of spin excitations is given by the imaginary part
the dynamic spin susceptibility which, within the present linear spin-wave
approximation, takes the following form:  
\be
{\rm Im} \chi_{{\bf q}}^{\gamma}(\omega)= 
\frac{\mid F_{\gamma}({\vc q}) \mid}{\omega_{\gamma}({\vc q})}\;
\delta {\Big(} \omega-\omega_\gamma(\vc q) {\Big)}.
\ee
In the paramagnetic phase, the factors $F_{\gamma}({\vc q})$ representing 
the spectral weights of $\gamma=x,y,z$ magnon modes are given by  
\bea
\bl
 &
 F_{x}({\vc q})= F_{y}({\vc q})=\frac{54{\bar \kappa_{2}}}{19} \phi_{\vc q};
 \;\; \;\;
 F_{z}({\vc q})=\frac{30{\bar \kappa_{1}}}{11} \phi_{\vc q}.
\el
\eea
In the magnetic phase I, we have 
\bea
\bl
 &
 F_{x}({\vc q})= F_{y}({\vc q})=\frac{27{\bar \kappa_{1}}}{11} \phi_{\vc q};
 \\
 &
 F_{z}({\vc q})=\frac{3{\bar \kappa_{1}}}{11} (-1+11\eta_{1})\phi_{\vc q} ,
\el
\eea
and, finally, for the magnetic phase II, we obtain 
\bea
\bl
 &
 F_{x}({\vc q})= \frac{3{\bar \kappa_{2}}}{19} (-1+19\eta_{2})\phi_{\vc q};
  \\
 &
 F_{y}({\vc q})=\frac{54{\bar \kappa_{2}}}{19} \phi_{\vc q};
 \\
 &
 F_{z}({\vc q})=\frac{60{\bar \kappa_{2}}}{19} \phi_{\vc q}.
\el
\eea
\begin{figure*}
\centerline{
\includegraphics[width=0.98\linewidth]{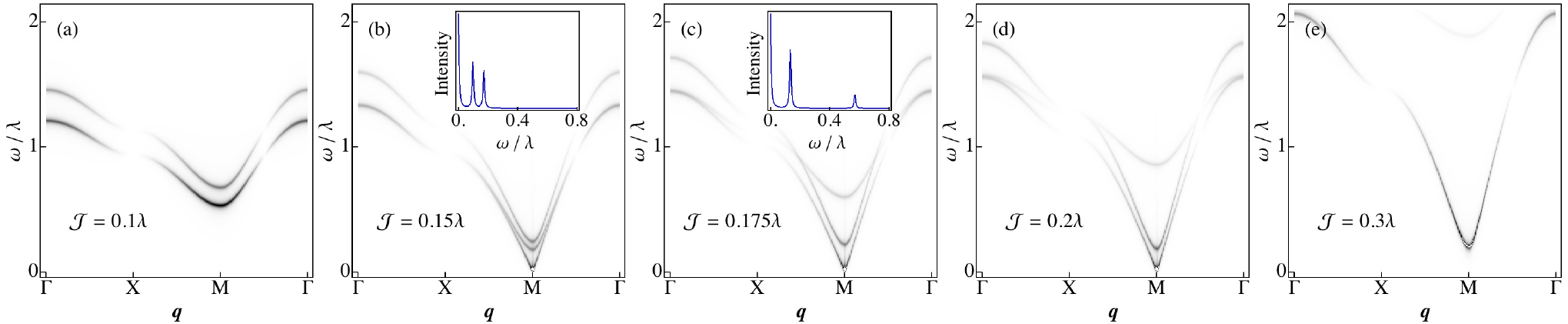}
}
\caption{ (Color online) 
Contour maps of the magnon intensities (multiplied by $\sqrt \omega$ 
for better visibility), $\sqrt \omega \;I({\bf q},\omega)=
\sqrt\omega \;{\rm Im}\chi_{{\bf q}}(\omega)$,   
for constant $\delta=0.2\lambda$ and different values of $\cal J$:  
(a) ${\cal J}=0.1\lambda$, (b) ${\cal J}=0.15\lambda$, 
(c) ${\cal J}=0.175\lambda$, (d) ${\cal J}=0.2\lambda$ and 
(e) ${\cal J}=0.3\lambda$. 
Note that (a) belongs to  paramagnetic phase, (b-d) represent the magnetic
phase II, and (e) represents the magnetic phase I [see the blue points 
in Fig.~\ref{MagnetizationFig}(c)]. The insets in (b) and (c) show a direct
comparison of the intensities $I({\bf Q},\omega)$ of three modes 
at the ordering wave-vector ($M$-point). The highest peak 
corresponds to the amplitude mode, the middle one represents out-of-plane
magnon, and the lowest peak is in-plane magnon (which is gapless hence 
not properly shown for numerical reasons).   
}
\label{Figsusceptibility}
\end{figure*}
Magnon intensities are given by 
$I({\bf q},\omega)=\sum_\gamma{\rm Im}\chi_{{\bf q}}^\gamma(\omega)$. The contour
plots of this quantity, multiplied by $\sqrt\omega$ for clarity, 
are shown in Fig.~\ref{Figsusceptibility}. In the $({\cal J}-\delta)$
parameter space, five different panels in this figure correspond to the blue 
points in Fig.~\ref{MagnetizationFig}(c), and thus represent (a) the
paramagnetic phase, (b-d) the magnetic phase II (${M  \parallel ab}$), and, 
finally, (e) the magnetic phase I (${M  \parallel c}$). 

In PM phase, the intensities of all (degenerate $x/y$, and $z$) modes are 
nearly equal. In phase II [panels (b-d)], which is of particular interest in 
the context of Ca$_2$RuO$_4$, the intensity of the highest energy (amplitude) 
mode is large near the critical point [see inset in panel (b)], but it fades 
away rather quickly at larger $\cal J$ values. 

\section{ Application   to ${\bf {\rm Ca}_2{\rm RuO}_4}$}

The calcium ruthenate, Ca$_2$RuO$_4$, has a layered perovskite structure 
similar to that of La$_2$CuO$_4$ cuprate, and shows a Mott-insulating behavior 
below room temperature~\cite{Nak97,Cao97}. It undergoes a magnetic phase 
transition at $\sim 110$~K, below which an antiferromagnetic order with a 
staggered moment $M\simeq 1.3\:\mu_\mathrm{B}$ is observed~\cite{Bra98}. 
A sizeable value of $LS$-product indicates that SOC is not 
quenched~\cite{Miz01}, and hence this material may exhibit some features 
of the ''excitonic'' magnetism considered above. To our knowledge, no 
dynamical spin susceptibility measurements for Ca$_2$RuO$_4$ have been 
reported to date; some theoretical expectations are given below.  

Observed $ab$-plane orientation of the moments~\cite{Bra98} is consistent with
the phase II in Fig.~\ref{PhaseFig}(b), which is stabilized by a compressive 
tetragonal distortion present in Ca$_2$RuO$_4$.  

One can roughly estimate the parameters $\cal J/\lambda$ and $\delta/\lambda$ 
from the observed staggered moment $M\simeq 1.3\:\mu_\mathrm{B}$~\cite{Bra98}
and the static magnetic susceptibility $\chi \simeq 2.6\times 10^{-3}
$emu/mol\cite{Bra98}. The moment $M$ is determined by $\eta_2$ [see
Eq.~(\ref{SMagnetization})] defining the distance to the critical point, 
while the susceptibility is given by  
\be
\chi_{ab}=\frac{12 \mu^2_B N_A}{\kappa_{2}(1+\eta_{2})}\;,
\ee
where $N_A$ is Avogadro number. From the $M$ and $\chi$ equations, we find  
$\eta_{2} \simeq 0.85$, and estimate the parameters 
${\cal J}/\lambda \sim 0.17$, and $\delta/\lambda \sim 0.2$. Magnon
dispersions in Ca$_2$RuO$_4$ are then expected to resemble the plots shown in 
Fig.~\ref{Figsusceptibility}(b,c). These plots suggest a full magnon bandwidth 
of the order of $1.5 \lambda\sim 100$~meV, given a spin-orbit coupling 
constant $\lambda (=\xi/2)\simeq 75$~meV~\cite{Miz01}. The parameter
${\cal J}=\frac{t_0^2}{U}\simeq 13$~meV which follows from these estimates
seems reasonable for $t_{2g}$ systems with $t_0\sim 0.2$~eV and $U\sim 3$~eV. 
As far as the amplitude mode is concerned, the insets in 
Fig.~\ref{Figsusceptibility} suggest a sizeable intensities; however, it might
be difficult to identify this mode because it falls in the phonon-energy window
($\sim 40$~meV). 

\section{Summary  }

We have studied here the phase diagram and magnetic excitations in Van
Vleck-type $d^4$ Mott insulators with spin-orbit singlet ground state. 
As the intersite exchange interactions increase, the system makes a
transition into an antiferromagnetically ordered state. For a square lattice 
geometry considered here, the exchange anisotropy supports a uniaxial-type 
magnetic order. Under a compressive strain, this order changes to the 
easy-plane one via a first order phase transition. We have calculated 
magnetic excitations over an entire phase diagram, quantifying the magnon 
dispersions and their intensities. We hope that the results presented here 
will motivate experimental studies of Ca$_2$RuO$_4$ and other potential 
candidate materials for excitonic-type magnetism~\cite{Cao14} by means 
of inelastic neutron and/or resonant x-ray scattering techniques. 

We would like to thank B.J. Kim for useful discussions.

\end{document}